\documentclass[12pt,preprint]{article}
\usepackage{epsf}
\newcommand{\eqb}{\begin{equation}}
\newcommand{\eqe}{\end{equation}}
\newcommand{\dmb}{\begin{displaymath}}
\newcommand{\dme}{\end{displaymath}}

\newcommand{\ep}{\varepsilon}
\newcommand{\eab}{\begin{eqnarray}}
\newcommand{\eae}{\end{eqnarray}}
\newcommand{\ra}{\right\rangle}
\newcommand{\la}{\left\langle}
\newcommand{\e}{\mbox{e}}
\newcommand{\be}{\begin{equation}}
\newcommand{\ee}{\end{equation}}

\begin{document}
\begin{titlepage}
\begin{flushright}
MPI-PhT 2001-?? \\
\end{flushright}
\vspace{0.6cm}

\begin{center}
\Large{{\bf Modifying operator product expansions by non-perturbative non-locality}}

\vspace{1cm}

Ralf Hofmann

\end{center}
\vspace{0.3cm}

\begin{center}
{\em Max-Planck-Institut f\"ur Physik\\ 
Werner-Heisenberg-Institut\\ 
F\"ohringer Ring 6, 80805 M\"unchen\\ 
Germany}
\end{center}
\vspace{0.5cm}

\begin{abstract}

Local quark-hadron duality violations in conventional applications of the operator product expansion 
are proposed to have their origin in the fact that the QCD vacuum or a hadronic state 
is not only characterized by 
nonvanishing expectation values of local, gauge invariant operators but also by finite 
correlation lengths of the corresponding gauge invariant $n$-point functions.
Utilizing high-resolution 
lattice information on these correlators a non-perturbative component of 
OPE coarse graining is proposed which, in principle, allows for a determination 
of the critical dimension where the break-down of the expansion sets in.

\end{abstract} 

\end{titlepage}

A practical and very successful approach to relate hadronic parameters 
like masses, decay constants and widths to a few universal 
parameters of the vacuum of Quantum 
Chromodynamics \cite{QCD} was introduced a long time ago \cite{SVZ}. 
The method of QCD sum rules (QSR's) relies on Wilson's operator product expansion (OPE) \cite{Wilson} at an 
external, euclidean momentum $Q$, analyticity of the considered QCD current correlator 
everywhere in the plane except for a cut along the
positive real axis, and perturbative calculability of the operator coefficients. 
non-perturbative effects are introduced as power corrections 
via the nonvanishing vacuum averages of local, gauge invariant operators. 
If the average of an operator of mass dimension $d$ is 
comparable with $\Lambda^{d}$, where $\Lambda$ is a typical hadronic scale 
then modulo numerical factors power corrections are suppressed by $(\Lambda/Q)^{d}$, and one may 
hope that the expansion approximates the correlator at least 
in an asymptotic sense \cite{Shifman}. To make contact with 
hadronic properties one assumes quark-hadron duality, namely the property of the 
dispersive part of the correlator to be described in 
terms of measured hadronic cross sections. By appropriately adjusting the external momentum 
QSR's relate the properties of the lowest resonance to the 
non-perturbative condensates. There is not much doubt that {\sl averaged} 
spectral functions relevant in QSR's are dual to the OPE \cite{Shifman}. 
In practice it is fair to say, however, that there are channels where the 
sum rule method seems to be jeopardized by abnormally 
large non-perturbative corrections (scalar, pseudoscalar).

However, it is painfully obvious that the conventional, practical OPE, 
which is typically truncated at dimension $d=6$ or $d=8$ does not account for 
pointwise or local duality (for a summary see the review articles \cite{Shifman,Bigi}). 
Usually the asymptotic nature of the OPE is blamed for this. Indeed, 
perturbatively calculated Wilson coefficients being of the form $(\alpha_s(Q_0)/(\alpha_s(Q))^\gamma\times (Q^2)^{-d}$, 
where $\gamma$ denotes an (effective) anomalous dimension and $Q_0$ is a normalization scale, 
do not allow for the observed ``wiggling'' of 
the spectral function obtained by taking the 
imaginary parts of a term-by-term continuation to time-like momenta. Since the expansion is 
truncated terms, 
which behave like $\exp[-\mbox{const}\sqrt{Q^2/\Lambda^2}]$, 
can not be captured \cite{Bigi}. On the other hand, such terms contribute asymptotically 
unsuppressed oscillations to the spectral function 
which contradicts experiment and is not allowed by asymptotic freedom. Investigations concerning 
the origin of local duality violation have been performed within various models in 
the literature \cite{Shifman0,Golterman,Shifman1}. Impressively, it was shown in \cite{Shifman0,Golterman} 
that upon appealing to a dispersion relation 
a model spectral function with equally spaced narrow resonances yields 
an asymptotic expansion in the euclidean which resembles a 
conventional OPE with factorially growing coefficients. However, such a model 
ignores the drastic broadening of 
resonances and the decrease of their heights 
with increasing energy.  

Local duality violations 
may have a considerable impact on the calculation of 
life-time differences and inclusive/exclusive decay widths 
of $B$-mesons where OPE in conjunction with 
the heavy quark expansion is applied at the large 
time-like momentum $p\sim m_b$ \cite{Nierste,Neubert}. The precise theoretical determination of these 
quantities is of particularily acute relevance since it 
would lead to stringent constraints on the parameters of the underlying model (e.g. CKM matrix 
elements in the Standard Model (SM) and/or couplings, masses, and CP violating phases of SM extensions). 
This, in turn, is of paramount importance for the potential detection of New Physics \cite{Nierste1}. 
It is therefore essential to improve our understanding of local duality and 
to estimate the scales which govern it.

The purpose of this 
letter is to point out and exploit a genuine source for OPE modification 
which has only been vaguely addressed so far \cite{Bigi}. The observation is that the conventional OPE 
of a current-current correlator, 
which can be proven to exist perturbatively, does 
not account for the diminishing {\sl relevance} of {\sl local operators} composed of the {\sl fundamental} 
fields in an asymptotically free theory like QCD when decreasing 
the spacetime resolution \cite{Hofmann}. This resolution, however, is determined by the external 
(euclidean) momentum squared $Q^2=-q^2$. 
The physical picture behind this can be examplified in the case of a 
correlator of light quark currents as follows \cite {SVZ}: An external current 
creates a quark-antiquark pair with total momentum $q$. 
Due to asymptotic freedom the 
propagation of these degrees of freedom is dominated by perturbation theory 
at high enough $Q\equiv\sqrt{Q^2}$ and comparable momenta. 
non-perturbative effects become important if there are soft 
emissions in the process, which are effectively described 
by nonvanishing vacuum expectation values (VEV's) of gauge invariant, {\sl local} operators. 
But what does {\sl local} really mean?
We believe that the answer to this question is tightly connected to the issue of 
pointwise duality. During the time interval 
between emission and reabsorption of the constituent degrees of 
freedom of a hadron the vacuum, which admits these propagations, is being 
resolved precisely by the external momentum $Q$. So if the OPE is an expansion involving 
{\sl local} operators one also ought to 
say how resolved the corresponding locality is. The renormalization group teaches that 
information on the VEV of some local operator at a 
given resolution derives from local and 
{\sl non-local} information at higher resolution. In perturbation theory, 
where interactions do not generate effective fields and parameters 
out of their fundamental counterparts,  
this is done by normalizing the 
corresponding operator at a definite momentum scale $Q_0$ 
and relating this to the momentum $Q$ of the process 
by means of the corresponding set of renormalization group equations. 
Since in QSR's the VEV of a local operator is 
supposed to {\sl exclusively} carry non-perturbative information one must apply an {\sl exclusively} 
non-perturbative procedure of coarse graining to it. 
The key to non-perturbative coarse graining for the lowest dimensional 
operators is information on the corresponding gauge invariant 2-point functions \cite{Dosch} 
which is available from lattice calculations 
\cite{DiGiacomo}. For higher dimensional operators one so far 
has to appropriately impose vacuum saturation to express their correlators in terms of the 
2-point functions. 

The idea that due to the existence of finite correlation lengths 
in gauge invariant 2-point functions non-local, non-perturbative effects should be included in the OPE is not new. 
We just mention here that to obtain 
realistic hadronic wave functions along the lines of ref.\,\cite{Zhit} 
non-local effects were systematically included via an expansion in covariant derivatives 
in ref.\,\cite{Mikhail}. On the other hand, based on perturbative arguments  
it was shown in refs.\,\cite{Karchev} that a 
light-cone expansion of the product of gauge
invariant currents can be written in terms of non-local string operators. Working in position space, perturbative 
evolution equations were derived for these operators in \cite{Braun}. 
In this paper we introduce non-perturbative, non-locality within a phenomenological generalization of the 
standard Wilson OPE. Although we 
concentrate on vacuum correlators it is clear that the procedure is 
extendable to hadronic state averages. Applications of the concept outlined above 
to OPE's and spectral functions of various light quark channels 
are reported in ref.\,\cite{PRD}.
 
We start out by expanding the correlator into 
a conventional OPE at a large external, euclidean momentum $Q$, where we expect the 
description of the dynamics in terms of the continuum action and local operators made of {\sl fundamental} fields 
to be sufficiently accurate. To proceed to lower external momenta Wilson coefficients are 
evolved perturbatively by means of the running coupling $\alpha_s(Q^2)$ and the 
anomalous dimensions of the operators. 
The evolution of an operator average can 
be obtained using information on the non-perturbative part of the corresponding 
gauge invariant correlator in euclidean position space
\eqb
\label{nl}
\la F(0)\ra_Q^{np} \equiv \la F_1(0)\cdots F_n(0)\ra_Q^{np} \to 1/{\cal N}\sum \la F_1(0)\cdots F_n(x_n)\ra_Q^{np} \ .
\eqe
Thereby, appropriate parallel transporters 
\eqb
S(0,x)\equiv{\cal P}\exp\left[ig\int_0^{x}dz_\mu\, A_\mu\right]\ .
\eqe
in the non-local expression are understood to make the 
object gauge invariant. The sum runs over all possible (piecewise straight \cite{Dosch}) trajectories of parallel transport. 
A normalization $1/{\cal N}$ depending on the mass dimension $d$ and the numbers of fields transforming 
under the fundamental and adjoint representation is implied  
to reduce the correlation function to the ``condensate'' in the limit $x_1,\cdots,x_n\to 0$. 
As examples we give the following dimension 5 and 6 operator VEV's 
\eqb
\label{exop}
(i)\ \la\bar{\psi}(0)F_{\mu\nu}(0)\sigma_{\mu\nu}\psi(0)\ra_Q^{np},\ 
\ \ \ \ \ \ \ \ \ (ii)\  \la \mbox{tr}\, F_{\mu\nu}(0)F_{\nu\kappa}(0)F_{\kappa\mu}(0)\ra_Q^{np}\ .
\eqe
If an arrow pointing from $x_i$ towards $x_j$ stands for 
the parallel transport $S(x_i,x_j)$ then the relevant contribution to the 
correlators correponding to the local operators of 
eq.\,(\ref{exop}) can be depicted as in Fig.\,1. Note that points with fields 
transforming under the fundamental (adjoint) representation are 
connected to one (two) lines of parallel transport. 
\begin{figure}
\vspace{4.3cm}
\includegraphics{Fig1.ps}
\caption{Diagrammatic representation of the gauge invariant correlators corresponding to 
the local operators of $(i)$ and $(ii)$ in eq.\,(\ref{exop}).} 
\label{} 
\end{figure}

To coarse grain from resolution $Q$ to resolution $Q-dQ$ we propose to average the non-perturbative part of a 
correlator corresponding to a {\sl connected} diagrams over a (euclidean) 
ball of radius $dR_Q$ with
\eqb
dR_Q=\frac{1}{Q-dQ}-\frac{1}{Q}=\frac{1}{Q}\left(\frac{1}{1-\frac{dQ}{Q}}-1\right)\sim \frac{dQ}{Q^2}\ .
\eqe
In the short-hand notation of eq.\,(\ref{nl}) this is written as
\eqb
\label{gc}
\la F(0)\ra_{Q-dQ}^{np}=\frac{1}{(V(dR_Q))^{(n-1)}}\int_{|x_1|,\cdots,|x_n|
\le dR_Q}d^4x_1\cdots d^4x_n \la F_1(0)\cdots F_n(x_n)\ra_Q^{np}\ ,
\eqe
where
\eqb
V(dR_Q)=\frac{1}{2}\pi^2 (dR_Q)^4=\frac{1}{2}\pi^2\,\left(\frac{dQ}{Q^2}\right)^4\ .
\eqe
As for correlators described by {\sl disconnected} diagrams the 
application of the factorization hypothesis (vacuum saturation) 
is discussed in \cite{PRD} for the example of 4-quark operators 
composed of color singlet or octet currents. 

Let us now concentrate on 2-point functions as they are relevant for dimension 3 and 4 
quark and gluon operators, respectively. 
To our knowledge only the gauge invariant bilocal quark correlators (vector and scalar) \cite{DiGiacomo} 
and the gluonic field 
strength correlator \cite{Dosch} have been measured on the lattice. The results 
suggest that there exists an additive decomposition into a perturbative, power-like in $|x|$, 
and a non-perturbative, exponential in $|x|$ piece \cite{DiGiacomo}\footnote{There have been a number of attempts 
to calculate gauge invariant quark and gluon correlators in the instanton vacuum \cite{Dor}. In this work, however, we do 
restrict ourselves to direct lattice results.} We are interested in the latter
\eqb 
\label{ex}
\la F_1(0)F_2(x)\ra_Q^{np}=A(Q)\,\exp(-|x|/\lambda)\ .
\eqe
Since $\lambda$ is expected to have direct physical meaning - see for example ref.\,\cite{Eidemuller}, where 
in the case of the field strength correlator $\lambda^{-1}$ was interpreted as 
the energy difference $\bar{\Lambda}$ between a heavy hybrid (an adjoint quark with glue) 
and the mass of the heavy quark - it can not depend on the resolution $Q$. 
So coarse graining may only affect the pre-exponential factor $A(Q)$. 
To derive an evolution equation for $A(Q)$ we combine eqs.\,(\ref{gc}) and (\ref{ex}):
\eqb
\label{gc2}
\left(V(dQ/Q^2)\right)^{-1}\,\int_{|x|\le \frac{dQ}{Q^2}}d^4x\, A(Q)\e^{-|x|/\lambda}=
A(Q-dQ)\e^{-0/\lambda}=A(Q-dQ)\ .
\eqe
Note that in the limit $\lambda\to\infty$ this leaves $A$ invariant. 
In this limit we recover the conventional treatment of non-perturbative corrections in the framework of the OPE. 
Expanding the l.h.s. and r.h.s. of eq.\,(\ref{gc2}) in $\frac{dQ}{Q^2}$ and comparing coefficients 
of the linear terms, we obtain
\eqb
\label{evo}
\frac{d}{dQ}\,A(Q)=\frac{4}{5\lambda}\,Q^{-2}A(Q)\ .
\eqe
Comparison of zero order terms yields an identity as it should. 
Evolution equations like eq.\,(\ref{evo}) could be derived for $(n>2)$-point functions 
if the explicit form of the non-perturbative part could be extracted from the lattice. 
The solution of (\ref{evo}) is 
\eqb
\label{run}
A(Q)=A(Q_0)\exp\left[-\frac{4}{5\lambda}\left(\frac{1}{Q}-\frac{1}{Q_0}\right)\right]\ .
\eqe
Disregarding powers of logarithms in the Wilson coefficients, which come from the 
anomalous dimensions of the corresponding operators and 
combined with the running coupling $\alpha_s(Q^2)$ possess 
exponents which are numerically small for the naively vacuum saturated 
4-quark operators in the OPE's of light-quark correlators \cite{SVZ}, 
eq.\,(\ref{run}) implies the following, generic form for the 
non-perturbative corrections of dimension 4 and 6 
\eqb
\label{pc}
\frac{A_4(Q_0)}{(Q^2)^2}\,\exp\left[-\frac{4}{5\lambda_4}\left(\frac{1}{Q}-\frac{1}{Q_0}\right)\right]\ ;\ \ \ 
\frac{A_6(Q_0)}{(Q^2)^3}\,\exp\left[-\frac{8}{5\lambda_6}\left(\frac{1}{Q}-\frac{1}{Q_0}\right)\right]\ .
\eqe
Here the term `naive' refers to vacuum saturation at 
the level of local operators and subsequent delocalization 
of the scalar 2-point functions (see also ref.\,\cite{PRD}). 
Eq.\,(\ref{run}) implies that 
conventional, dimension 4 and 6 power corrections are drastically 
altered if $Q$ is of the order of $\lambda^{-1}$ or less. Despite the fact that the expression for 
dimension 6 in eq.\,(\ref{pc}) heavily relies on the vacuum 
saturation hypothesis it may feature a general trend of decreasing (effective) correlation lengths in 
higher mass dimensions (ultimately, only the lattice can decide on the truth of 
this assertion).

Let us extrapolate 
the behavior of eq.\,(\ref{pc}) to non-perturbative corrections 
of arbitrary dimension $d$. 
If the running with $Q$ essentially can be expressed as 
\eqb
\label{pca}
\frac{A_d(Q_0)}{Q^d}\,\exp\left[-\lambda^{-1}_d \left(\frac{1}{Q}-\frac{1}{Q_0}\right)\right]\ ,
\eqe
where $\lambda_d$ denotes an effective correlation length, 
then there is a single maximum at momentum $Q=(d\,\lambda_d)^{-1}$ with value
\eqb
P_d\equiv\left(\Lambda_d\,\lambda_d\,\exp[-1]\,d\right)^d \,\exp\left[\frac{\lambda^{-1}_d}{Q_0}\right]\ ,
\eqe
where $\Lambda_d\equiv (A_d(Q_0))^{(1/d)}$. 
So a non-perturbative correction at {\sl fixed} $d$ vanishes asymptotically for both, 
$Q\to\infty$ and $Q\to 0$, which is in stark 
contrast to the conventional picture. This, however, 
does by no means imply that we approach 
asymptotic freedom at low momenta since higher and higher mass 
dimensions may contribute their maxima at lower and lower $Q$. 
One possibility for this to happen is the following: Let $\lambda_d$ fall off 
as $\lambda_d=\lambda/d^{(1-\ep)}$, $(1>\ep>0)$, then the positions of the 
maxima still decrease as $Q=\lambda^{-1} d^{-\ep}$. The values of these 
maxima behave as  
\eqb
P_d=\left(\Lambda_d\,\lambda\, d^\ep\,\exp[-1+\frac{d^{-\ep}}{\lambda Q_0}]\right)^d\ . 
\eqe
In particular, assuming that $\Lambda_d\equiv \Lambda_{QCD}$ as 
it is suggested by conventional OPE's, the critical 
dimension $d_c$ from which on maxima explode is given implicitely as 
a function of $\lambda$, $\ep$, $\Lambda_{QCD}$, and $Q_0$ by 
\eqb
\Lambda_{QCD}\,\lambda\, d^\ep\,\exp[-1+\frac{d^{-\ep}}{\lambda Q_0}]=1\ .
\eqe
Truncating the 
OPE at some $d_t<d_c$ would mean that physics 
below the corresponding $Q$ is not described. 
A truncation at $d_t\ge d_c$ does not make 
sense since due to the rapid increase of the contributions at $d\ge d_c$ 
we are surely at orders where the expansion does not approximate (asymptotic expansion). 
Let us give some numerical examples. For $\ep=0.5$, $\Lambda_{QCD}=0.4$ GeV we calculate 
(the rounded) $d_c$ as a function of $\lambda$ 
\eab
\lambda&=&3\,\mbox{GeV}^{-1}\ \ \rightarrow\ \ d_c=4\ ;\ \ \ \ \ \ \,
\lambda=2\,\mbox{GeV}^{-1}\ \ \ \ \ \rightarrow\ \ \ d_c=10\ ;\nonumber\\ 
\lambda&=&1\,\mbox{GeV}^{-1}\ \ \rightarrow\ \ d_c=40\ ;\ \ \ \ \ 
\lambda=0.5\,\mbox{GeV}^{-1}\ \ \ \rightarrow\ \ \ d_c=158\ . 
\eae
Lattice measurements with $N_F=4$ staggered fermions of mass $a\cdot m_q=0.01$ and 
lattice resolution $Q_0=a^{-1}\sim 2$ GeV suggest that fermionic and 
gluonic correlation lengths corresponding to scalar correlation functions 
are $\sim$ 3.2 GeV$^{-1}$ and $\sim 1.7$ GeV$^{-1}$, respectively \cite{DiGiacomo}. 
So for dimension 4 and assuming (naive) vacuum saturation 
also for dimension 6 this does 
not pose a problem for the conventional 
sum rule analysis of light quark correlators (at 
$Q_0=a^{-1}\sim 2$ GeV $A_{\bar{q}q}(Q_0)$ as well as $A_{F^2}(Q_0)$ seem to be 
compatible with their phenomenological 
values at $\mu\sim 1$ GeV \cite{DiGiacomo}). 
However, (naive) vacuum saturation and/or the exclusion of higher mass dimensions do 
not lead to a realistic behavior of the {\sl spectral functions} which is shown in 
\cite{PRD}. Still, going only to dimension 6, 
one obtains a much more realistic picture if the (effective) 
correlation length at this dimension is drastically decreased (by about 10 times!). 
This can be motivated by the small correlation length 
of the 2-point vector function 
on the one hand (naive vacuum saturation not applicable). 
On the other hand this short correlation 
length may cumulatively simulate higher mass dimensions. The tendency that dimension 6 contributions 
are overestimated in the conventional OPE of \cite{SVZ} for distances $>0.3$\,fm can be extracted 
from the work of \cite{SS}. There, the euclidean position space $V\pm A$ correlators were 
calculated in a random instanton liquid model and 
compared to the ALEPH data \cite{Aleph} and the conventional OPE. 
Note, however, that fitting experimental data to a 
conventional OPE does violate local duality. 
We investigate the issue in \cite{PLB}.

In the modern literature there is supporting evidence in 
favor of large mass scales (as compared to $\Lambda_{QCD}\sim 0.2-0.4$ GeV) 
characterizing the QCD vacuum: Based on lattice data for the instanton size distribution it was stated 
in ref.\,{Shuryak} that the average size of instantons 
is much smaller ($\sim 1/3$ fm) than their mean separation ($\sim 1$ fm). 
Using maximal Abelian projection on the lattice to identify magnetic monopoles, 
the size of these defects, $r\sim  0.05$ fm, 
was determined for an SU(2) pure gauge theory ($\beta\sim 2.5$) in ref.\,\cite{Born}. 
To calibrate the lattice the SU(3) value of the 
string tension $\sigma=440$ MeV was used. 
It was found in \cite{CNZ} that a phenomenologically motivated 
tachyonic gluon mass can take values well above the 1 GeV bound.

To summarize, we proposed to use non-local, non-perturbative information 
on the QCD vacuum (extendable to hadronic states) for a non-perturbative 
procedure to coarse grain the euclidean VEV's (or hadronic matrix elements) 
of local operators composed of fundamental fields down to resolutions 
where their relevance must be questioned. Implications for the low-energy behavior of 
the OPE in the euclidean were discussed. In particular, a possibility for the determination of the 
critical operator dimension, where the breakdown of the OPE sets in, 
was outlined.

\section*{Acknowledgements}    

The author would like to thank Uli Nierste for a stimulating conversation. Financial support 
from CERN's theory group during a research stay in June are gratefully acknowledged. 
The author is indebted to V. I. Zakharov for numerous useful discussions and valuable comments.

\bibliographystyle{prsty}

\end{document}